\documentclass[12pt,preprint]{aastex}

\newcommand{\be}{\begin{equation}}
\newcommand{\ee}{\end{equation}}
\newcommand{\ben}{\begin{eqnarray}}
\newcommand{\een}{\end{eqnarray}}
\newcommand{\bc}{\begin{center}}
\newcommand{\ec}{\end{center}}

\begin{document}
   \title{Gravitational lensing as a possible explanation for some
   unidentified gamma-ray sources at high latitudes}

\author{Diego F. Torres\footnote{Physics Department, Princeton University, NJ 08544. E-mail: {\em
dtorres@princeton.edu} }, Gustavo E. Romero\footnote{Instituto
Argentino de Radioastronom\'{\i}a, C.C.5, 1894 Villa Elisa, Buenos
Aires, Argentina. Member of CONICET. E-mail: {\em
romero@irma.iar.unlp.edu.ar} }, and Ernesto F. Eiroa
\footnote{Instituto de Astronom\'{\i}a y F\'{\i}sica del Espacio,
C.C. 67, Suc. 28, 1428, Buenos Aires, Argentina. E-mail: {\em
eiroa@iafe.uba.ar} }}


\begin{abstract}
   We propose that some of the high-latitude unidentified EGRET
$\gamma$-ray sources could be the result of gravitational lensing
amplification of the innermost regions of distant, faint, active
galactic nuclei. These objects have $\gamma$-ray emitting regions
small enough as to be affected by microlensing of stars in
interposed galaxies. We compute the gravitational amplification
taking into account effects of the host galaxy of the lens and
prove that, whereas the innermost $\gamma$-ray regions can be
magnified up to thousand times, there is no amplification at radio
frequencies, leading to the observed absence of strong
counterparts. Some new effects in the spectral evolution of a
gravitational microlensed $\gamma$-ray AGN are predicted. Within a
reasonable range of lensing parameters, and/or types of sources,
both variable and non-variable EGRET detections at high latitudes
can be explained by microlensing. The same phenomenon could also
have an important incidence among the future GLAST detections at
high-latitudes.

\end{abstract}

\section{Introduction}

The Third EGRET Catalog of $\gamma$-ray sources includes
observations carried out between April 22, 1991 and October 3,
1995, and lists 271 point-like detections (Hartman et al. 1999).
About two thirds of them have no conclusive counterparts at lower
frequencies, and are dubbed unidentified.

These unidentified $\gamma$-ray sources can be divided in two
broad groups. The first one, at low latitudes, is probably related
to several galactic populations such as radio-quiet pulsars,
interacting supernova remnants, early-type stars with strong
stellar winds, X-ray binaries, etc (see Romero 2001 and references
therein). The second group of unidentified sources is formed by
mid- and high-latitude detections. Gehrels et al. (2000) have
shown --in a model independent way-- that the mid-latitude sources
are different from the bright population of unidentified sources
along the galactic plane. Some of the mid-latitude detections
($5^{\circ}<|b|<30^{\circ}$) are thought to be associated with the
Gould Belt (Grenier 2000, Gehrels et al. 2000), a starburst region
lying at $\sim600$ pc from Earth. A few other sources, at higher
latitudes, might be the result of electrons being accelerated at
the shock waves of forming clusters of galaxies (Totani et al.
2001). However, for most of the high-latitude unidentified sources
(HL-UnidS), no other explanation seems to be available than they
are AGNs yet unnoticed at lower energies (Reimer \& Thompson
2001).

A $\gamma$-ray AGN population consisting of 66 members has already
been detected by EGRET (Hartman et al. 1999). Photon spectra and
variability indices of the HL-UnidS match well those of known
$\gamma$-ray AGNs. In Figure 1, we show the corresponding
distributions, with the AGN-histogram adapted from Torres et al.
(2001a). The variability criterion used here is the $I$-scheme,
but other variability criteria yield similar results (Torres et
al. 2001b). Population studies have already remarked that part of
the sample of HL-UnidS is consistent with an isotropic population,
a fact that also supports an extragalactic origin (\"Ozel \&
Thompson 1996).

All identified $\gamma$-ray AGNs are also strong radio sources
with flat spectrum, as expected from synchrotron jet-like sources
where the $\gamma$-ray flux is the result of inverse Compton
scattering (Mattox et al. 1997). We could ask, then, why the
HL-UnidS are not detected at lower frequencies, particularly in
the radio band, if they are also AGNs? Here we propose that some
of these sources are the result of gravitational lensing
amplification of background, high-redshift, active galactic
nuclei; blazars whose $\gamma$-ray emitting regions are small
enough as to be affected by microlensing by stars in interposed
galaxies.

\section{Gravitational lensing of active galactic nuclei}

Since AGNs have emission regions of different size for different
wavelengths, we expect a differential magnification of light. The
innermost regions of AGNs, responsible for the $\gamma$-ray
emission, have linear sizes $x \sim 10^{14}$--10$^{15}$ cm (e.g.
Blandford and Levinson 1995). Significant gravitational lensing
requires that $D_{{\rm ol}}/D_{{\rm os}} \times x/2 < R_E$, where
${D_{{\rm ol}}}$ and ${D_{{\rm os}}}$ are the angular-diameter
distances between the lens (l), or the source (s), and the
observer (o), and $R_E$ is the Einstein radius of the lens. For
typical redshifts (say, $z\sim$ few tenths for the lens, and
$z\sim 0.8-3$ for the source), in a standard cosmological model,
we find that stars of few solar masses have Einstein radii
$\sim$500 times bigger than the projected AGN's $\gamma$-ray
emitting region, and thus, that the latter can be perfectly
amplified. At the same time, since radio emission is originated
far down the jet, the sizes of the corresponding radio regions
($x>10^{17}$ cm) exceed the Einstein radius of the lenses and lead
to the absence of radio counterparts. In the case of the optical
emission, if it is the same particle population giving rise -in
the same region- to gamma-rays (through inverse Compton
scattering) and optical emission (through synchrotron radiation),
then we can also expect gravitational magnification of the optical
luminosity. In some cases with high (lensing) optical depth, we
could even expect simultaneous intensity variations in both bands,
something that could in principle be tested with the improved
capabilities and source location accuracy of the forthcoming GLAST
satellite. Optical monitoring of stellar-like sources inside GLAST
error boxes could lead to the identification of new gamma-ray
blazars, and to some knowledge of their redshift distribution. One
interesting fact is that the spectral evolution that we predict
below will only happen in the gamma-ray band, since it critically
depends on the size of the gamma-ray spheres (given by the opacity
to pair production), a process not operative in the optical band.
The idea that some extreme properties of distant AGNs can be the
result of gravitational microlensing is not new (e.g. Ostriker \&
Vietri 1985, Gopal-Krishna \& Subramanian 1991, Romero et al.
1995). However, microlensing effects upon the $\gamma$-ray
emission have not been discussed in detail yet.

It is usually assumed that the a priori probability of finding a
small group of distant, gravitationally magnified objects is below
1\%. Indeed, recent results (E.L. Turner, personal communication,
2001) taking into account the clustering of lenses in interposed
galaxies, give values between 10$^{-2}$ -- 10$^{-3}$ for the a
priori probability of finding gravitational magnified sources in
random directions of the sky. In addition, for those directions
where there is gravitational lensing, the probability of it
happening with optical depth above 0.2 is very high. In our case,
the number of potential compact $\gamma$-ray emitting background
sources is extremely large: only the last version of the
V\'eron-Cetty \& V\'eron's (2001) Catalog--which is still very
incomplete at high redshifts-- contains more than $10^3$ already
identified blazars, in addition to more than $10^4$ quasars and
other less energetic AGNs. Actually, GLAST mission enhanced
sensitivity is expected to uncover more than $10^4$ $\gamma$-ray
emitting AGNs (Gehrels \& Michelson 1999). If the actual number of
$\gamma$-ray emitting AGNs below the EGRET detection threshold is,
say, 10$^7$ (which can be a conservative assumption: approximately
1 $\gamma$-ray blazar per $10^4$ galaxies), they might produce
high-latitude sources lacking a clear low-frequency counterpart.
Even when considering reduced probabilities for microlensing with
large magnifications (i.e. $\sim \tau/A^2$, where $\tau$ is the
local optical depth and $A$ the magnification), we still found
that a handful of the sources at high latitudes already observed
by EGRET, and maybe a few hundreds of sources to be observed by
GLAST, can be the result of microlensing. For instance,
considering a random lensing probability of 5 10$^{-3}$, and a
local value of $\tau$ of ${\cal O} (1)$ (even larger values are
possible because the presence of superluminal components in
blazars --see below--) and amplifications of $\sim 100$, we could
produce by lensing $\sim 5$ high-latitude sources (as we see
below) without counterparts at lower energies. Of course, if a
large number of distant blazars are, say, one order of magnitude
below GLAST sensitivity, in such a way that they do not need huge
amplifications in order to be detected, the number of possible
microlensing cases could very much increase. Assuming an
amplification of a factor 10, the number of cases could increase
to several hundreds in the forthcoming large GLAST sample.

\section{Microlensing model for extended sources}

We consider a distant and weak $\gamma$-ray emitting blazar whose
GeV flux is well below EGRET sensitivity threshold and whose radio
flux is at the mJy level, also below the sensitivity of typical
all-sky surveys. Assuming a galaxy interposed in the line of sight, the
lens equation, in the lens plane, is (Chang \& Refsdal 1979,
Schneider et al. 1992)
\begin{equation}
{\bf r}-{\bf r}_{0} -R_{E}^{2}\frac{{\bf r}}{r^{2}}-
{\bf diag} (\kappa +\gamma, \kappa -\gamma )
{\bf  r}-{\bf  \omega}_{0}=0,  \label{p1}
\end{equation}
where the coordinate system is centered on the lens, with the
orientation of the orthonormal basis $\{{\bf  e}_{1},{\bf
e}_{2}\}$ chosen to diagonalize the quadrupole matrix; the source
is at ${\bf  r}_{0} $ and the image position is ${\bf  r}$. The
fourth and fifth terms in Eq. (\ref{p1}) arise from the deflection
imprinted by the host galaxy as a whole: $\kappa $ and $\gamma $
are the focusing and the shear of the galaxy at the point mass
position, respectively.  The only effect of ${\bf  \omega }_{0}$
is to change the unperturbed source position ${\bf  r}_{0}$ by a
constant. We shall ignore $ {\bf  \omega }_{0}$ assuming a source
position ${\bf  s}={\bf  r}_{0}+ {\bf  \omega }_{0}$ in the lens
plane. Defining new appropriate coordinates in the lens and the
source plane (Schneider et al. 1992),
${\bf  X}=[\sqrt{\left| 1-\kappa +\gamma \right| }/R_{E}]{\bf r}$,
and ${\bf  Y}=[1/R_E\sqrt{\left| 1-\kappa +\gamma \right|}]{\bf
s}$, the lens equation becomes ${\bf  Y}=\varepsilon \;\left(
\begin{array}{cc}
{\bf diag} (\Lambda, 1)
\end{array}
\right) {\bf  X}-{\bf  X}/\left| {\bf  X}\right| ^{2} \label{p5}
$, where $\varepsilon =sign\left( 1-\kappa +\gamma \right)$ and $
\Lambda =(1-\kappa -\gamma ) /( 1-\kappa +\gamma )$. The solution
of the latter equation can be found reducing the problem to a
fourth order equation for $X^{2}$ (Schneider et al. 1992).
We finally obtain the magnification, $A=I_{{\rm obs}}/I_{0}$, as
\be A=\frac {\left| 1-\kappa +\gamma \right|^{-1} \left(
X_{1}^{2}+X_{2}^{2}\right) ^{2}}{ \left| \Lambda \left(
X_{1}^{2}+X_{2}^{2}\right) ^{2}+\varepsilon \left( 1-\Lambda
\right) \left( X_{1}^{2}-X_{2}^{2}\right) -1\right| }. \label{p12}
\ee For extended circular sources, the magnification is given by
(e.g. Han et al. 2000):
\begin{equation}
A=\frac{\int_{0}^{2\pi }\int_{0}^{r_{\rm s}}{\mathcal I}(r,\varphi
)A_{0}(r,\varphi )rdrd\varphi }{\int_{0}^{2\pi
}\int_{0}^{r_{\rm s}}{\mathcal I} (r,\varphi )rdrd\varphi },
\label{e1}
\end{equation}
where $(r,\varphi )$ are polar coordinates in a reference frame
placed at the center of the source, $r_{\rm s}$ is the radius of
the source, ${\mathcal I }(r,\varphi )$ is the surface intensity
distribution of the source, and $A_0$ is the sum of the
magnifications of all images. We define $R=r/R_{E}$ and suppose
that the lens is moving with constant velocity ${\bf
v}$.\footnote{We choose the origin of time ($t=0$) as the instant
of closest approach between the lens and the source. Then, if the
center of the source is placed at ${\bf  b}=(b_{1},b_{2})$ when
$t=0$ (projected on the plane of the lens), the position of any
point of the source with polar coordinates $(r,\varphi )$ is $
s_{1}(t)=b_{1}-vt\cos \theta +r\cos \varphi $ and $
s_{2}(t)=b_{2}-vt\sin \theta +r\sin \varphi $, where $v=\left|
{\bf  v}\right| $ , $\theta $ is the angle between $ {\bf  v}$ and
${\bf  e}_{1}$, $0\leq r\leq r_{s}$ and $0\leq \varphi \leq 2\pi
$. In units of the Einstein radius, $ Y_{1}=[B_{1}-T\cos \theta
+R\cos \varphi]/ \sqrt{\left| 1-\kappa +\gamma \right|}$, and
$Y_{2}=[B_{2}-T\sin \theta +R\sin \varphi] /\sqrt{| 1-\kappa
+\gamma | }$, where $T=vt/R_{E}$ and $B_{1,2}=b_{1,2}/R_{E}$. When
$\gamma =0$ (no shear),  we can take $\theta =0$ and $ {\bf
B}=(0,B_{0})$ without loosing generality.}

The radius of the internal regions of AGNs depends on the energy
as a power law, $ r_{\gamma }\propto E^{p }$ (Blandford and
Levinson 1995). We then define a reference source with radius
$r_{{\rm ref}}$ and $\gamma$-ray   energy $%
E_{{\rm ref}}$ such that $ R_{\gamma }(E)=R_{{\rm ref}} (E/E_{{\rm
ref}})^{p } $, where capital letters stand for normalized
quantities using the Einstein radius. We assume that the intensity
of the source is uniform, and that its spectrum follows
approximately a power law $I_{0}(E)=I_{{\rm ref}} (E/E_{{\rm
ref}})^{-\xi } $ with $\xi \in (1.7,2.7)$, where $I_{{\rm ref}}$
is the intensity of the reference source (Krolik 1999). The
surface intensity distribution of the source will then be ${
\mathcal I}_{0}(E)= I_{0}(E)/ \pi R_{\gamma}(E)^{2}$, from where
the magnification can be computed.

Using that $A=I/I_{0}$, we define $ J\equiv I/I_{{\rm ref}}=A (
E/E_{{\rm ref}} ) ^{-\xi }$, the intensity in units of $I_{{\rm
ref}}$. We adopt $p =1.1$ , $\xi =2$, and a reference source with
dimensionless radius $R_{{\rm ref}}=2\times 10^{-3}$ and
$\gamma$-ray energy $E_{{\rm ref}}=0.1$ GeV (Blandford and
Levinson 1995). In Figure 2 we show just an example of the results
of our numerical computations. The $\gamma$-ray emission of the
background AGN can be amplified significantly. This magnification
can make an otherwise unnoticed source to exceed the detection
threshold. Note also that the $\gamma$-sphere corresponding to 10
GeV, whose size is similar to the optical emitting regions of some
typical AGNs, is negligibly amplified, while the lower energy
curves --well within the EGRET range-- all show magnifications in
excess of a factor 100. This phenomenon also has a particular
spectral signature, produced by the differential amplification of
the different $\gamma$-regions. The spectral evolution
(chromaticity) effect is a change in the spectral slope at medium
energies. This break, predicted only for microlensing events, and
its peculiar time evolution --it shifts towards high energies as
times goes by since transit-- can be used as a specific test to
differentiate this from other phenomena (see Figure 2, right
panels).

\section{Time scales and number of events}


The time scale of lensing variability is $t_0=R_E/v$. For
instance: for $M=0.1 M_\odot$, $H_0$=75 km s$^{-1}$ Mpc$^{-1}$,
$z_{\rm s}=1$, $z_{\rm l}=0.1$, $\Omega_0$=0.2, and $v$=5000 km
s$^{-1}$, $t_0 \sim 170$ days, while the half width of the peaks
shown in Figure 2 is about 9 days. This may result in a variable
source for separated EGRET viewing periods (typically of $\sim 15$
days each). If, instead, $M_{{\rm l}} \sim 5 M_\odot$ and $v \sim
1000 $ km s$^{-1}$, the half width of the peaks is $\sim $1000
days, and the $\gamma$-ray source would most likely be seen as a
steady, non-variable detection. \footnote{The velocities chosen
are typical for extragalactic objects. For galaxies at redshifts
$z\sim 0.1$ or bigger, their recession velocities very much exceed
the one we are using (by more than order of magnitude). The latter
are relative velocities, involving small traversal components of
the recession speed and proper motions of the star within the
galaxy and the galaxy itself (with respect to its cluster). A
worked example  (for BL LAC 0846+51) can be found in Nottale
(1986), for which he quotes $300<v<3000$ km/s.}

Following Subramanian \& Gopal-Krishna (1991), if the lenses have
the same mass, and the observing period is $\Delta t$, the
expected number of events, in the case of a background source
moving with velocity $v$, will be \be N=\tau \left( 1+ \frac{2
\Delta t}{\pi u t_0} \right),
\end{equation}
where $\tau $ is the optical depth and $u=b/r_E=B/R_E$  is usually
taken as 1. The second term between the brackets corresponds to
the increment of the probability of detecting a microlensing event
due to the movement of the source (see Romero et al. 1995 and
Surpi et al. 1996). The optical depth can be defined as the ratio
of the surface mass density of microlensing matter to the critical
mass density $\Sigma_{{\rm crit}}= c^2 D_{\rm os}/4\pi G D_{\rm
ol} D_{\rm ls}$. For simplicity, we shall assume that the lenses
are located at a distance from the center of the galaxy much
smaller than the core radius. Then, $\kappa \sim \Sigma_{{\rm
c}}/\Sigma_{{\rm crit}}$, with $\Sigma_{{\rm c}}$ the central
surface density. The high surface mass density associated with the
core of normal galaxies along with the usual assumption that most
of this mass is in the form of compact objects naturally leads to
high optical depths for microlensing. For instance, in the case of
the lensed quasar Q2237-031, where four images are well-resolved,
lensing models indicate values of $\tau\sim0.5$ (Schneider et al
1988), which are corroborated by the detections of
microlensing-based optical variability with relatively high duty
cycles (e.g. Corrigan et al. 1991). Other lensed sources display
even higher duty cycles (e.g. Koopmans \& de Bruyn 2000).

For a mass distribution of lenses given by $ N(M)\propto M^{-\alpha}$ for
$M_{\min}\leq M\leq M_{\max}$ (Salpeter 1955), the number of expected
microlensing events by stars with masses in the range $(M_{1}, M_{2})$,
included in the total mass range $(M_{\min}, M_{\max})$, during $\Delta t$
days of observations is \begin{equation} N_{M_{1}-M_{2}}^{\Delta
t}=2.16\sqrt{\kappa\tau}\frac{v}{c}
\frac{D(\alpha,M_{1},M_{2})}{B(\alpha,M_{\min},M_{\max})}\frac{\Delta
t}{30{\rm days}} \end{equation} where the functions $B$ and $D$ are,
respectively, \begin{eqnarray} B(\alpha,M_{\min},M_{\max})&=& \frac{\left[
\left( \frac{M_{\max}}{M_{\odot}}\right) ^{2-\alpha}-\left(
\frac{M_{\min}}{M_{\odot}}\right) ^{2-\alpha}\right] }{2-\alpha},\nonumber
\\ &&\;\;\;\;\;\; {\rm for}\;\; \alpha\neq2, \nonumber \\ &=& \ln\left(
\frac{M_{\max}}{M_{\min}}\right)  , \;\;{\rm for}\;\; \alpha=2,
\end{eqnarray} and \begin{eqnarray} D(\alpha,M_{1},M_{2}) &=&
\frac{2\left[ \left( \frac{M_{2}}{M_{\odot}}\right)
^{\frac32-\alpha}-\left( \frac{M_{1}}{M_{\odot}}\right) ^{\frac32
-\alpha}\right]}{3-2\alpha},\nonumber \\ &&\;\;\;\;\;\;\;\; {\rm
for}\;\;\alpha\neq\frac{3}{2}, \nonumber \\ &=& \ln\left(
\frac{M_{2}}{M_{1}}\right), \;\;{\rm for}\;\; \alpha=\frac{3}{2},
\end{eqnarray} and we have assumed a case with $z_{\rm s}\sim 0.9$ and
$z_{\rm l}\sim0.1$ to fix the numerical coefficient. Clearly, the total
number of events will strongly depend (apart from the expected influence
of $\tau$ and $\kappa$) on the velocity of the source and the index
$\alpha$, usually taken in the range $(2, \;3)$.

A particular interesting case of source candidates are blazars
where the high-energy emission is produced in a superluminal
component with apparent velocity $v>c$ in the lens plane
(Gopal-Krishna \& Subramanian 1991). They will produce
$\gamma$-ray sources with the highest levels of variability. We
find that during the EGRET's lifetime, hundreds of events can
be expected for optical depths in the range 0.2--0.4. Even for
values of $\tau$ as low as $10^{-3}$, if the source is apparently
superluminal, then the number of expected events in the EGRET
observing time is above 10. Instead, sources whose velocities in
the lens plane are much smaller than $c$ (say, about 5000 km
s$^{-1}$) will produce only a few events. A complete study of the
distribution of lens masses within this model will be presented
elsewhere.




\section{Concluding remarks}

In summary, we have found that gravitational microlensing of the
innermost regions of distant AGNs can produce unidentified sources
compatible with those observed at high galactic latitudes. While
large amplification factors are obtained for $\gamma$-rays, a
negligible magnification results in the radio band. In the case of
the optical emission, if it is the same particle population giving
rise to gamma-rays and optical emission then we can also expect
magnification of the optical luminosity. The gamma-ray spectral
evolution provides a specific signature for the microlensing
events that can be used to differentiate this from other kind of
phenomena. Higher (lensing) optical depth, or the presence of
shear, will lead to a diversity of light curves. We remark that we
have used the Chang and Refsdal (1979) model as the gravitational
lensing scenario, but that galaxies with denser cores would
require a more detailed treatment. This would necessarily include
the study of caustic patterns of all stars at a time in order to
get the light curves. We shall explore the lensing model in much
more detail in
a subsequent publication (Eiroa et al. 2002, in preparation).\\

\acknowledgements

This work has been partially supported by UBA (UBACYT X-143, EFE),
CONICET (DFT, and PIP 0430/98, GER), ANPCT (PICT 98 No. 03-04881,
GER), and Fundaci\'{o}n Antorchas (separate grants to GER and
DFT). GER thanks the kind support from the astrophysics group at
the MPIfK, Heidelberg, where part of his research was carried out.
Prof. Turner is acknowledged for providing his -at that time-
unpublished results, and for useful discussions. Drs. R.C.
Hartman,  D.J. Thompson and S. Ritz are acknowledged for comments
and discussions. We are especially grateful to Dr. Hartman for his
insightful remarks.

\clearpage

\begin{figure}
\plottwo{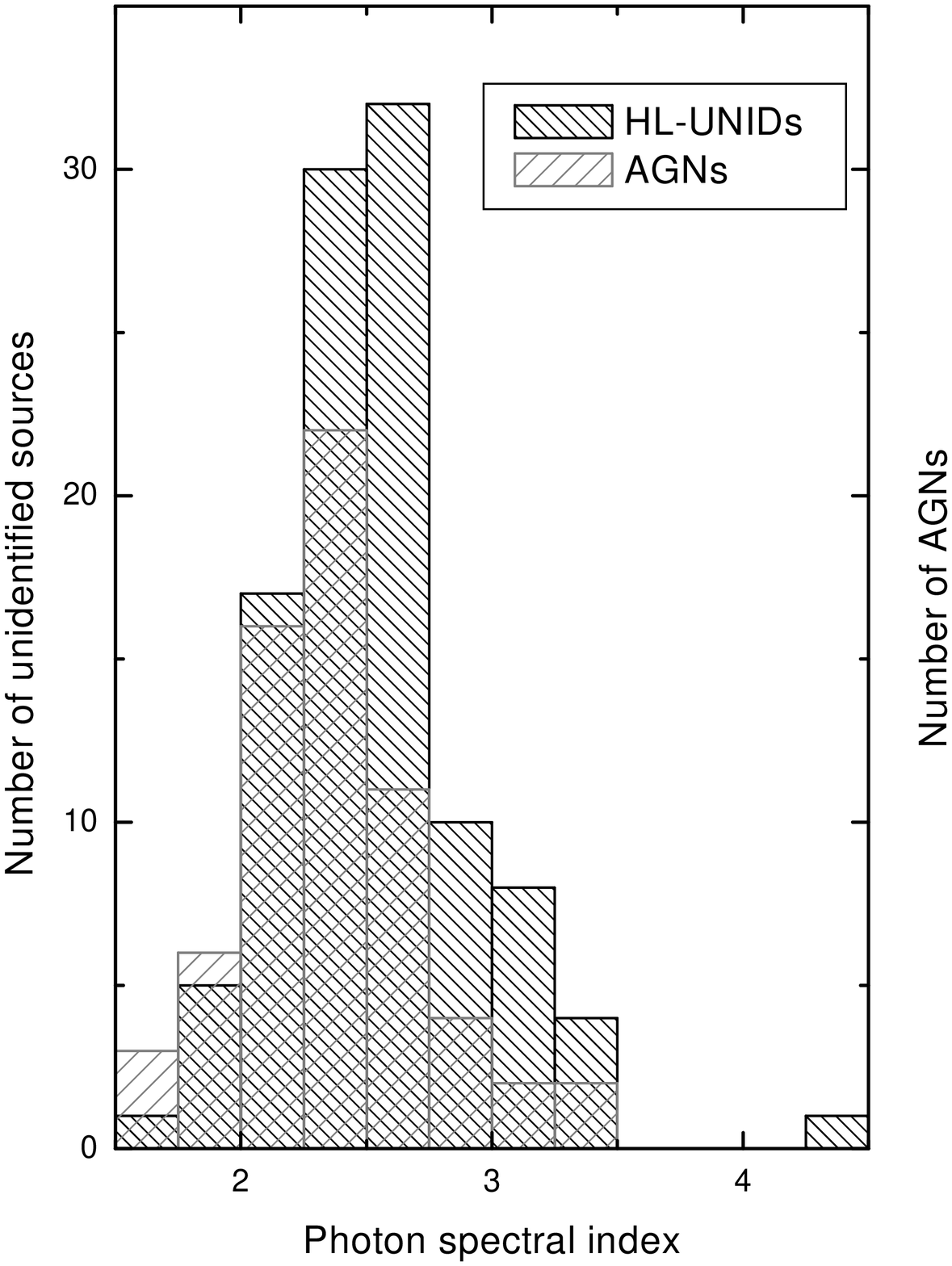}{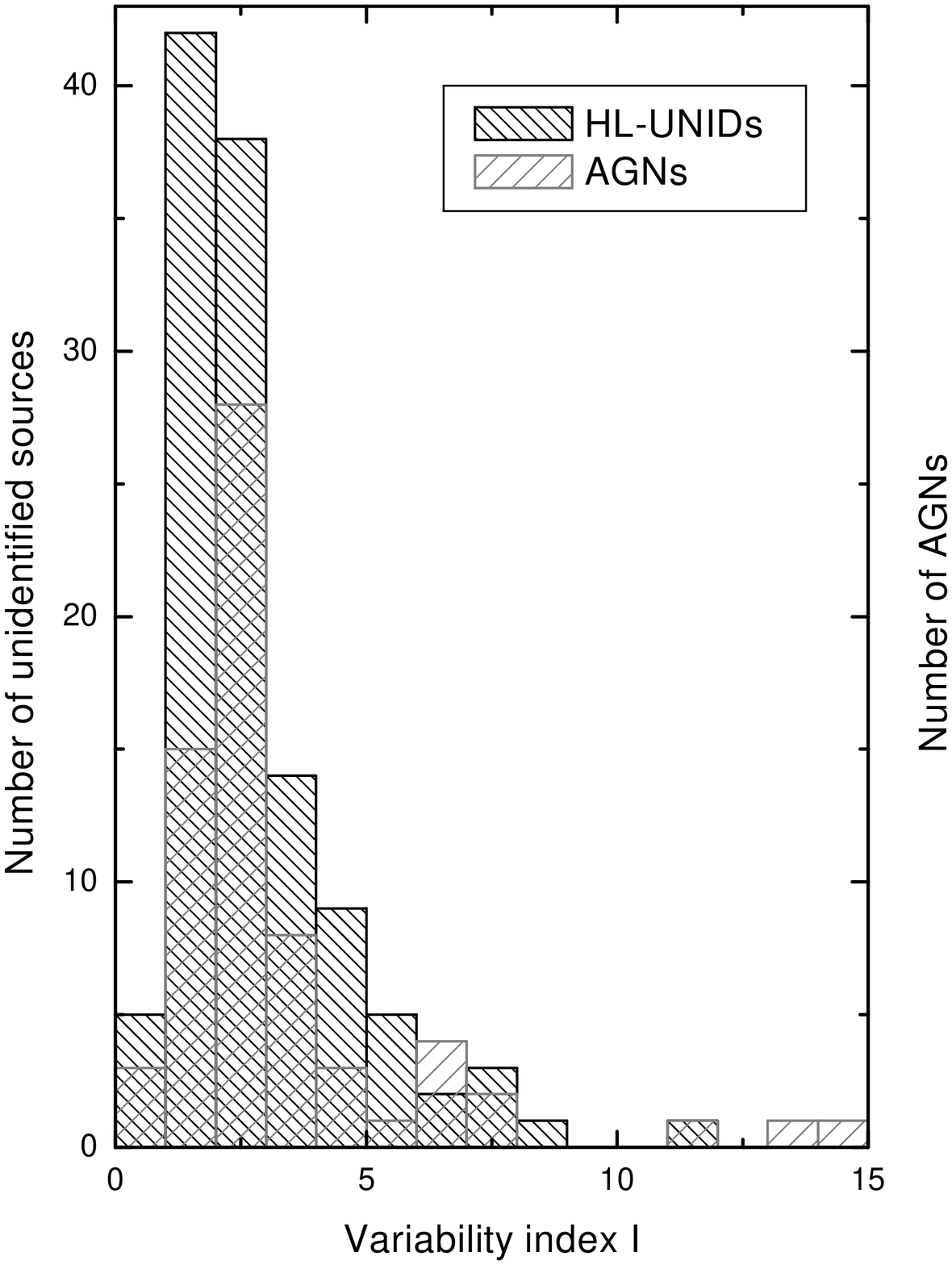} \caption{Left: Distribution of the photon
spectral index of all unidentified sources with galactic latitudes
$|b|>10$ deg. A Gaussian fit would give a mean equal to 2.6 and
deviation equal to 0.3. Most of the sources have steep spectra,
consistent with the AGN population. Right: Histogram for the
variability index $I$ of each of the HL-UnidS ($I=1$ is the mean
value of $I$ for pulsars). A Gaussian curve would have in this
case a mean at $I$=2, a value already 2$\sigma$ above the expected
one for a non-variable population, and a deviation equal to 0.7.
More than half of these sources are classified as likely variable.
Taking only those sources having $|b|>40^{\circ}$, both mean
values are even higher. See Torres et al. (2001a,b) for details.
}\label{fig1}
\end{figure}

   \begin{figure}
\epsscale{.8} \plotone{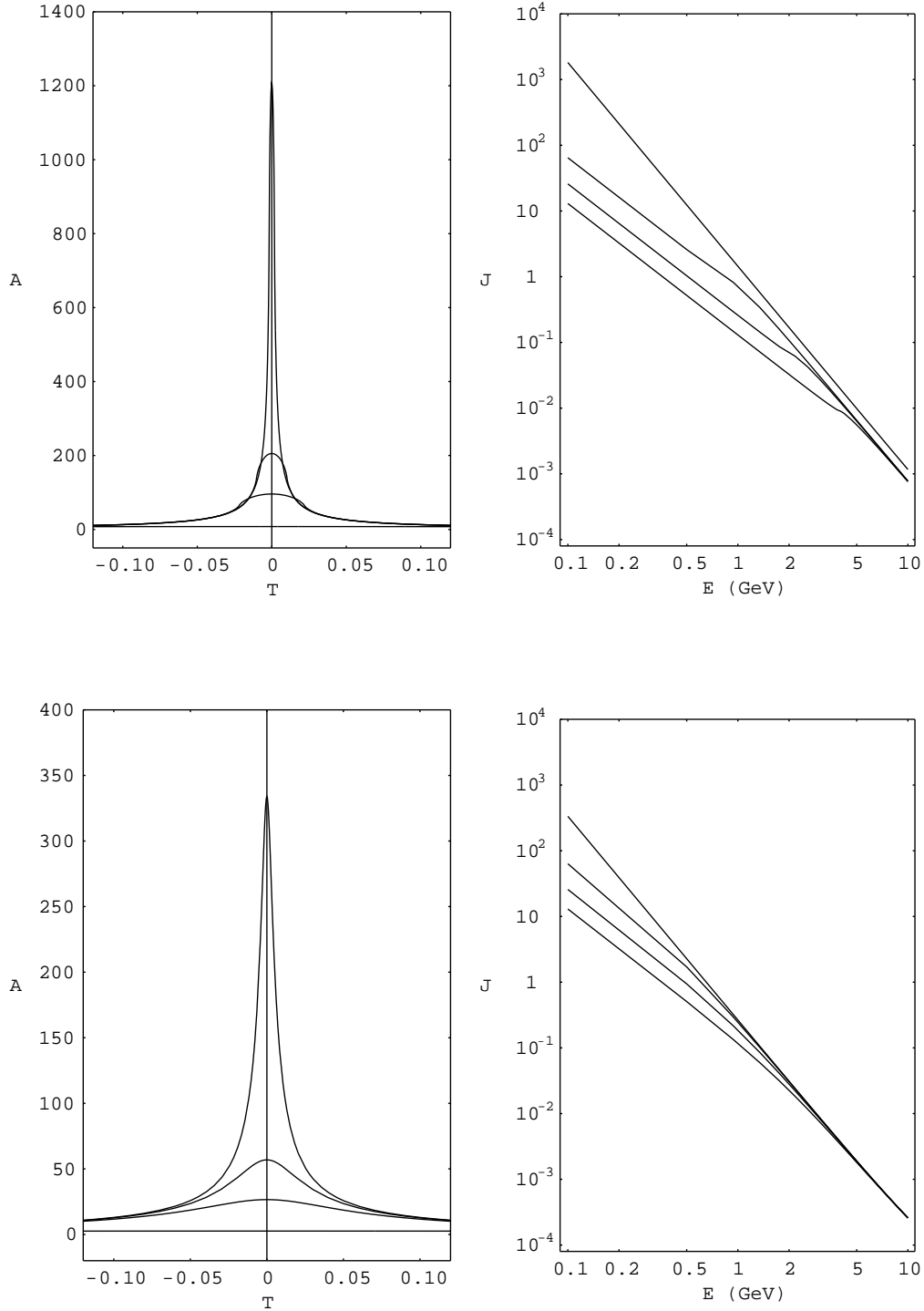}   \caption{Lensing results for a
dimensionless impact parameter equal to $B/R_s=b/r_s= 0.5$ (upper
panel) and 2 (lower panel). Notice the specific feature (a
changing break in the power-law spectrum) predicted for the
spectral evolution. In the left panels we show the light curves,
from top to bottom E= 0.1 GeV, 0.5 GeV, 1 GeV, 10 GeV. In the
right ones the spectrum evolution, from top to bottom
$T=vt/R_E=$0, 0.02, 0.05, and 0.10. Lensing parameters were chosen
as $\kappa=0.4$, $\gamma=0$.
 } \label{fig:nlens}
\end{figure}

\end{document}